\documentclass[12pt,preprint]{aastex}

\slugcomment{Accepted for publication in the Astrophysical Journal}

\let\url\relax
\usepackage[bookmarks=true,bookmarksnumbered=true,colorlinks,citecolor=blue,linkcolor=blue,urlcolor=blue]{hyperref}
\hypersetup{
  pdfauthor = {Gutierrez, K. et~al.},
  pdftitle = {Multiwavelength Observations of 1ES 1959+650, One Year After 
the Strong Outburst of 2002},
  pdfsubject = {Astrophysics},
  pdfcreator = {LaTeX with hyperref package}
}
\shortauthors{K.\ Gutierrez eat al.}
\shortauthors{K.\ Gutierrez et al.}
\shorttitle{1ES1959+650}
\begin{document}
\title{Multiwavelength Observations of 1ES 1959+650, One Year After 
the Strong Outburst of 2002}
\author{K.~Gutierrez,\altaffilmark{1}
              H.~M.~Badran,\altaffilmark{2}
	      S.~M.~Bradbury,\altaffilmark{3}
	      J.~H.~Buckley,\altaffilmark{1}
	      O.~Celik,\altaffilmark{4}
              Y.~C.~Chow,\altaffilmark{4}
              P.~Cogan,\altaffilmark{5}
	      W.~Cui,\altaffilmark{6}
	      M.~Daniel,\altaffilmark{5}
	      A.~Falcone,\altaffilmark{6}
	      S.~J.~Fegan,\altaffilmark{4}
	      J.~P.~Finley,\altaffilmark{6}
	      G.~H.~Gillanders,\altaffilmark{5}
              J.~Grube,\altaffilmark{3}
	      J.~Holder,\altaffilmark{3}
	      D.~Horan,\altaffilmark{7}
	      S.~B.~Hughes,\altaffilmark{1}
              I.~Jung,\altaffilmark{1}
	      D.~Kieda,\altaffilmark{8}
	      K.~Kosack,\altaffilmark{1}
	      H.~Krawczynski,\altaffilmark{1}
	      F.~Krennrich,\altaffilmark{9}
	      M.~J.~Lang,\altaffilmark{5}
	      S.~Le~Bohec,\altaffilmark{8}
	      G.~Maier,\altaffilmark{3}
	      P.~Moriarty,\altaffilmark{10}
	      J.~Perkins,\altaffilmark{1}
	      M.~Pohl,\altaffilmark{9}
	      J.~Quinn,\altaffilmark{5}
              P.~F.~Rebillot,\altaffilmark{1} 
	      H.~J.~Rose,\altaffilmark{3}
	      M.~Schroedter,\altaffilmark{4,7}
	      G.~H.~Sembroski,\altaffilmark{6}
	      S.~P.~Wakely,\altaffilmark{11}
	      T.~C.~Weekes,\altaffilmark{7}
	      R.~J.~White,\altaffilmark{3} (The VERITAS Collaboration)}
\and 
\author{M.~Aller,\altaffilmark{12} H.~Aller,\altaffilmark{12} P.~Charlot,\altaffilmark{13} J.~F.~Le~Campion\altaffilmark{13}}
 
              \altaffiltext{1}{Department of Physics, Washington University, 
		St. Louis, MO 63130, USA}  
	      \altaffiltext{2}{Physics Department, Tanta University,
		Tanta, Egypt}
	      \altaffiltext{3}{School of Physics and Astronomy, University of
		Leeds, Leeds, LS2 9JT, Yorkshire, England, UK}
	      \altaffiltext{4}{Department of Physics, University of
		California, Los Angeles, CA 90095-1562, USA}
	      \altaffiltext{5}{Department of Physics, National
		University of Ireland, Galway, Ireland}
	      \altaffiltext{6}{Department of Physics, Purdue
		University, West Lafayette, IN 47907, USA}
	      \altaffiltext{7}{Fred Lawrence Whipple Observatory,
		Harvard-Smithsonian CfA, P.O. Box 97, Amado, AZ 85645-0097} 
	      \altaffiltext{8}{High Energy Astrophysics Institute,
		University of Utah, Salt Lake City, UT 84112, USA}
	      \altaffiltext{9}{Department of Physics and Astronomy,
		Iowa State University, Ames, IA 50011-3160, USA}
	      \altaffiltext{10}{School of Science, Galway-Mayo
		Institute of Technology, Galway, Ireland}
	      \altaffiltext{11}{Enrico Fermi Institute, University of
		Chicago, Chicago, IL 60637, USA}
	      \altaffiltext{12}{University of Michigan, Ann Arbor, MI 48109, 
		USA}
              \altaffiltext{13}{Observatoire de Bordeaux (OASU) 
               -- CNRS/UMR 5804, BP 89, 33270 Floirac, France}
\email{Corresponding authors: Kris Gutierrez <mrground@hbar.wustl.edu> and Henric Krawczynski <krawcz@wuphys.wustl.edu>}
\begin{abstract}
In April-May 2003, the blazar 1ES 1959+650 showed an increased 
level of X-ray activity. This prompted a multiwavelength 
observation campaign with the Whipple 10 m $\gamma$-ray telescope,
the {\it Rossi X-ray Timing Explorer}, the Bordeaux Optical Observatory, and
the University of Michigan Radio Astrophysical Observatory.
We present the multiwavelength data taken from May 2, 2003 to 
June 7, 2003 and compare the source characteristics with those 
measured during observations taken during the years 2000 and 2002.
The X-ray observations gave a data set with high signal-to-noise
light curves and energy spectra; however, the $\gamma$-ray observations did not
reveal a major TeV $\gamma$-ray flare. Furthermore, we find that the radio and 
optical fluxes do not show statistically significant deviations from 
those measured during the 2002 flaring periods.
While the X-ray flux and X-ray photon index appear
correlated during subsequent observations, the  apparent correlation
evolved significantly between the years 2000, 2002, and 2003.
We discuss the implications of this finding for the mechanism 
that causes the flaring activity.
\end{abstract}
\section{Introduction}
\label{Gutierrez:intro}
Blazars are a class of Active Galactic Nuclei (AGN) with collimated plasma 
outflows (jets) directed along the line of sight.
The jets give rise to a continuum emission extending from
the radio to X-rays, sometimes even into the MeV and GeV/TeV energy range.
The blazar 1ES 1959+650 is one of ten blazars detected so far in the
GeV/TeV energy range with ground based Cherenkov telescopes \citep{Kraw:05,Tave:04}.
With a redshift $z\,=$ 0.047 \citep{Scha:93}, it is further away from us 
than the two strongest TeV $\gamma$-ray sources Mrk 421 
($z\,=$ 0.031) and Mrk 501
($z\,=$ 0.034) but substantially closer than the recently detected TeV emitters
H~2356-309 ($z\,=$ 0.165) and 1ES 1101-232 ($z\,=$ 0.186).

In the year 2002, 1ES 1959+650 went into a state of high TeV $\gamma$-ray
activity \citep{Hold:03,Ahar:03} and was observed intensively with good 
multiwavelength coverage (\citet{Kraw:04}, called ``Paper I'' 
in the following). The source showed very hard X-ray energy spectra 
with 3-25 keV photon indices $\Gamma$ ($dN/dE\propto E^{-\Gamma}$)  between 
1.6 and 2.4 (Paper I). The fact that its spectral energy 
distribution peaks sometimes at energies well above 15 keV 
makes the source one of the most extreme synchrotron 
blazars, similar to Mrk 501 \citep{Pian:98}
and 1H 1426+428 \citep{Falc:04}.

The time-averaged TeV $\gamma$-ray energy spectrum from the flaring period
of 2002
has been fitted with a power law model, giving a photon index 
$\Gamma$ of 2.8 over the energy range 
from 316 GeV to 10 TeV \citep{Ahar:03,Dani:05}.
Although the TeV $\gamma$-ray spectrum of 1ES 1959+650
is steeper than that of Mrk 501 
in its high state ($\Gamma\approx 2.2$) \citep{Ahar:99}, the difference may be entirely 
caused by the larger redshift of the source and hence the larger extent
of extragalactic absorption in pair production processes of the
TeV $\gamma$-rays with photons from the Cosmic Infrared Background (CIB)
\citep{Schr:05}.

Interestingly, the source showed an ``orphan'' flare (a
flare in the TeV $\gamma$-ray band without a corresponding flare in 
the X-ray band) on June 4, 2002. While the TeV flux increased from 0.26$\pm$0.21 
times the flux to the Crab Nebula to about 4 times the flux of
the Crab Nebula within 5 hrs, the X-ray flux, X-ray photon index, and
the optical brightness stayed approximately constant.
This 2002 observation challenges simple one-zone Synchrotron Self-Compton (SSC) models
where the X-rays originate as synchrotron emission of a single relativistic 
electron population and 
the TeV $\gamma$-rays are produced from inverse Compton scattering of the
synchrotron photons by the same population of electrons.
Paper I discusses several possible explanations for the orphan flare:
multiple emission zones contributing to the observed radiation,
time variable external seed photons fields for inverse Compton processes,  
an ordered magnetic field aligned with the jet axis, and
hadronic rather than leptonic emission models.
In a more recent paper, \citet{Bott:05} studied the 
possibility that the orphan flare originated from relativistic 
protons ($\gamma_{\rm P}\,=10^3-10^4$) in $p\,\gamma\,\rightarrow$
$\Delta^+\,\rightarrow$ $p\,\pi_0$ and $\pi_0\rightarrow$ $\gamma\,\gamma$
processes inside the jet as the protons interacted with jet photons
backscattered into the jet by a cloud several parsecs 
from the central engine. He finds that reasonable variations in the
values of the model parameters explain all 
the data satisfactorily.
A recent twist in the story of the orphan flare is its possible 
association with a $\sim$TeV neutrino detected by the AMANDA neutrino 
telescope from the direction of 1ES 1959+650 in temporal
coincidence, however, the statistical significance
cannot be reliably estimated \citep{Halz:05}. If 1ES 1959+650 indeed emitted 
high energy neutrinos, purely leptonic synchrotron-Compton 
models would be ruled out.

In this paper we report on multiwavelength observations of 1ES 1959+650
performed between May 2, 2003 and June 7, 2003. During the week preceding 
our observations, the {\it All Sky Monitor} on board of the {\it Rossi X-ray 
Timing Explorer (RXTE)} satellite measured an elevated 2-12 keV flux 
on the order of 10 mCrab. We therefore invoked pointed {\it RXTE} observations 
and accompanying observations with the Whipple 10 m Cherenkov telescope, 
the Bordeaux optical telescope, and the University of Michigan 
Radio Astrophysical Observatory (UMRAO).
While the TeV $\gamma$-ray observations did not reveal any major TeV $\gamma$-ray
flare and the radio and optical data did not show significant 
flux variability, we did acquire an X-ray data set with light curves 
and energy spectra with high signal-to-noise ratio.
We will discuss the data sets in Section~\ref{Gutierrez:Data}, and present the
multiwavelength light curves in Section~\ref{Gutierrez:Results}.
We study the secular evolution, 
i.e.~long term evolution, of the X-ray/TeV $\gamma$-ray flux correlation and
the X-ray flux versus X-ray spectral hardness correlation in 
Section~\ref{Gutierrez:Corr}. 
We study the spectral energy density (SED) and fit
it with a SSC model in Section~\ref{Gutierrez:SED}.
We summarize and discuss the results in Section~\ref{Gutierrez:discussion}, 
focusing on a very interesting result from the X-ray observations: a  
secular evolution of the possible 
X-ray flux versus X-ray photon index correlation. Throughout the
paper we will compare results from 2003 with those obtained in 2002 (Paper I)
and in earlier observations in 2000 \citep{Gieb:02}.
In the following, fit results are given in the text and in the tables 
with 90\% confidence interval errors; in figures, we show 1 $\sigma$ 
error bars and 90\% confidence level upper limits computed
with the method described in \cite{Hele:83}.

\section{Data Sets and Data Reduction}
\label{Gutierrez:Data}

Observations were taken with the University of Michigan Radio 
Astrophysical Observatory (UMRAO) between May 2 and June 29, 2003,
with the Bordeaux optical telescope between June 7 and October 18, 2003, 
and with the X-ray satellite {\it RXTE} and the Whipple 10~m Cherenkov 
telescope between May 2 and June 7, 2003. 
Our X-ray observations added 34.5~ks of integration time to the 
2000 data from \citet{Gieb:02} and the 2002 data from Paper I which 
we reanalyzed here. The TeV $\gamma$-ray data comprise 13.1 hrs.
Detailed descriptions of the data cleaning and analysis procedures 
are given in Paper I and in \citet{Char:04}.

For the X-ray analysis, we use here the \verb+FTOOLS+~v5.3.1 package for
data cleaning and the \verb+Sherpa+~v.3.0.1 package for spectral fitting.
Power law and broken power law models were fitted over the energy range
from 4 to 15 keV, using a galactic neutral hydrogen column density of 
1.01~$\rm\times~10^{21}$~cm$^{-2}$ \footnote{This value was
obtained at \url{http://heasarc.gsfc.nasa.gov/cgi-bin/Tools/w3nh/w3nh.pl}} 
We exclude here the 3-4 keV data, as past experience has shown that
it has a somewhat lower reliability than the 4-15 keV data.
We limit the analysis of individual {\it RXTE} pointings to $<$15 keV
(rather than to $<$25 keV as in Paper I), as the 15-25 keV band
is dominated by Poissonian background fluctuations for 
the ``low flux'' data sets from 2000 and 2003.

We quote TeV fluxes as integral fluxes in units of the flux from the Crab Nebula.
Given the zenith angle range of the observations, the peak 
energy\footnote{The peak energy is defined as the energy at which the differential $\gamma$-ray 
detection rate peaks, assuming a source with the same 
spectrum as the Crab Nebula.} of the flux measurements lies at 600 GeV.
Based on the Whipple measurements of the energy spectrum from the Crab 
Nebula \citep{Hill:98}, a flux of 1 Crab corresponds to a 1 TeV 
$\nu F_\nu$ flux of (5.12$\pm$0.27$_{\rm stat}$$\pm$$0.96_{\rm syst}$) 
$\times 10^{-11}$ ergs cm$^{-2}$ s$^{-1}$. 

\section{Light Curves}
\label{Gutierrez:Results}

\begin{figure}
\plotone{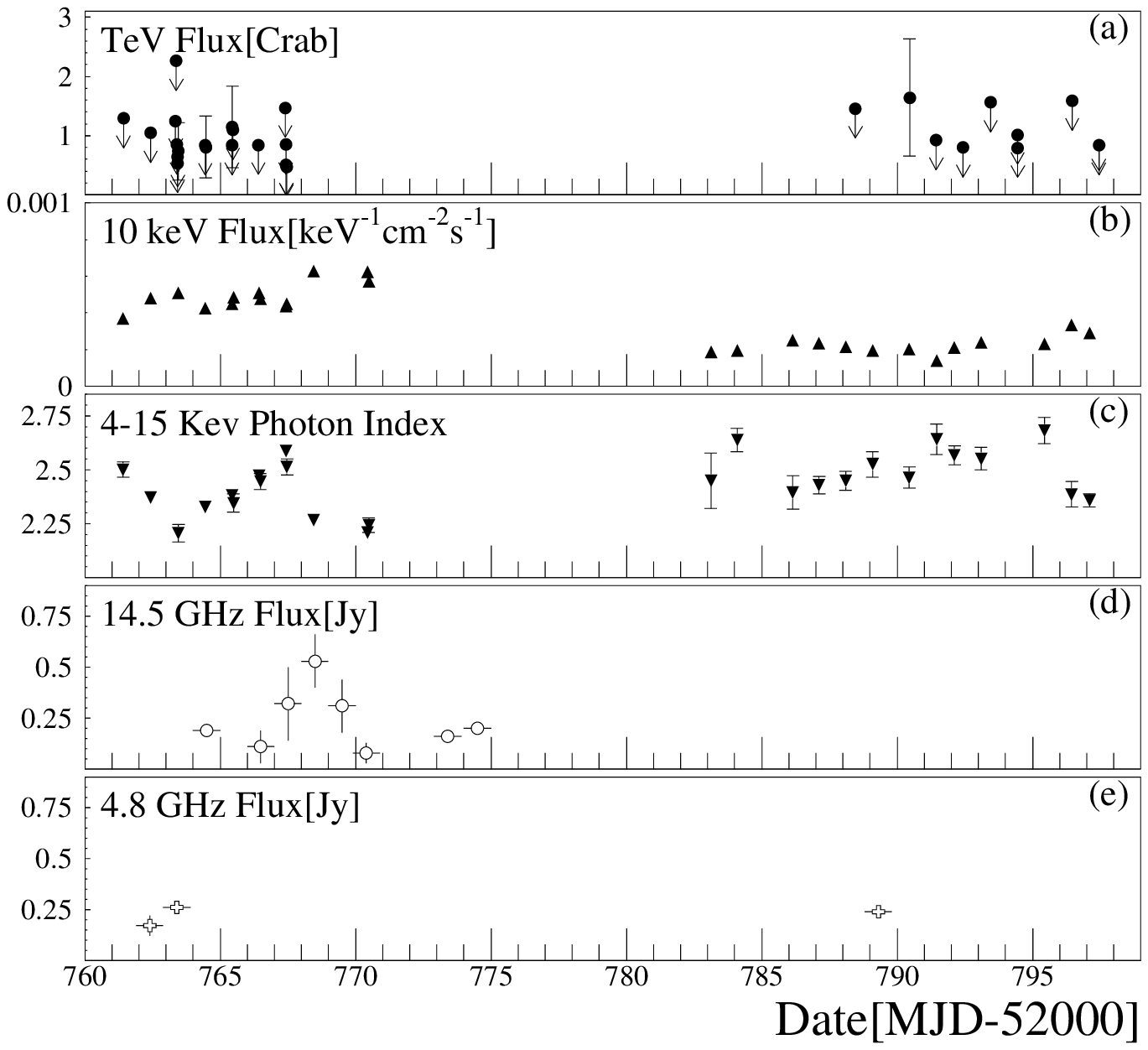}
\caption{Results from the 1ES 1959+650 multiwavelength campaign (May 2, 2003 -- June 7, 2003):
(a) TeV $\gamma$-ray flux (Whipple) in Crab units; the data are binned in 28 min bins; 
(b) the 10 keV X-ray flux (RXTE);
(c) the 4-15 keV X-ray photon index (RXTE);
(d) the 14.5~GHz flux density (UMRAO);
(e) the 4.8 GHz flux density (UMRAO).}
\label{fig1}
\end{figure}

We briefly discuss here the light curves proceeding from short wavelengths to long wavelengths.
The TeV $\gamma$-ray data (Figure~\ref{fig1}a) shows mostly upper limits. 
The observations did not reveal strong flares. Analyzing the entire TeV 
$\gamma$-ray data set, the significance of the $\gamma$-ray signal over the 2 months
observation was 3.3 $\sigma$ with a $\gamma$-ray rate of $0.24\pm0.11$ Crab units.  
Thus even on the longest accessible time scales, we have only marginal evidence
for TeV $\gamma$-ray emission from the source. The large ``gap'' in the $\gamma$-ray light 
curve corresponds to the full moon period when Cherenkov telescopes cannot be 
operated in standard mode and some additional down-time owing to bad weather.\\[2ex]
The 10~keV X-ray flux (Figure~\ref{fig1}b) varies by a factor of 3.4 over 
the course of the observations. 
The X-ray flux seems to be correlated with the 4--15~keV photon 
index $\Gamma$ (Figure~\ref{fig1}c) in the sense that higher fluxes are 
accompanied by harder energy spectra.\\[2ex]
We do not show the optical data in Figure~\ref{fig1} as only one observation 
was taken during the multiwavelength campaign. During this one observation, 
the V-band optical magnitude was $15.27\pm0.05$~mag.  The majority of optical 
observations were taken after the multiwavelength campaign. The flux measured during 
54 nights between Jun.~7 and Oct.~18, 2003, varied between 14.89~mag and 15.67~mag with 
an average of $15.21\pm0.05$~mag. During the 2002 flaring phase, the V-band 
magnitudes varied between 15.4~mag and 15.7 mag (Paper I).
The combined 2002 and 2003 data thus show no evidence for correlation between the
major X-ray/TeV $\gamma$-ray flaring phases and emission levels in the optical.
The 14.5~GHz and 4.8~GHz radio data are shown in Figs.~\ref{fig1}d and \ref{fig1}e respectively. 
We do not show here the 8~GHz data as it consists only of two data points.
The radio data do not show significant evidence for flux variability and the observed 
values are consistent with those measured earlier (Paper I, \citet{Greg:91,Beck:91}). 
The data show that the fractional flux variability of 1ES 1959+650 in the radio band 
is 10\% or less. Radio observations with higher sensitivity are needed to access 
any correlation between the radio fluxes and the X-ray or TeV $\gamma$-ray fluxes.
\\[2ex]

\section{Secular Evolution of the X-Ray/TeV Flux Correlation and the X-Ray Flux/Spectrum Correlation}
\label{Gutierrez:Corr}

\begin{figure}
\plotone{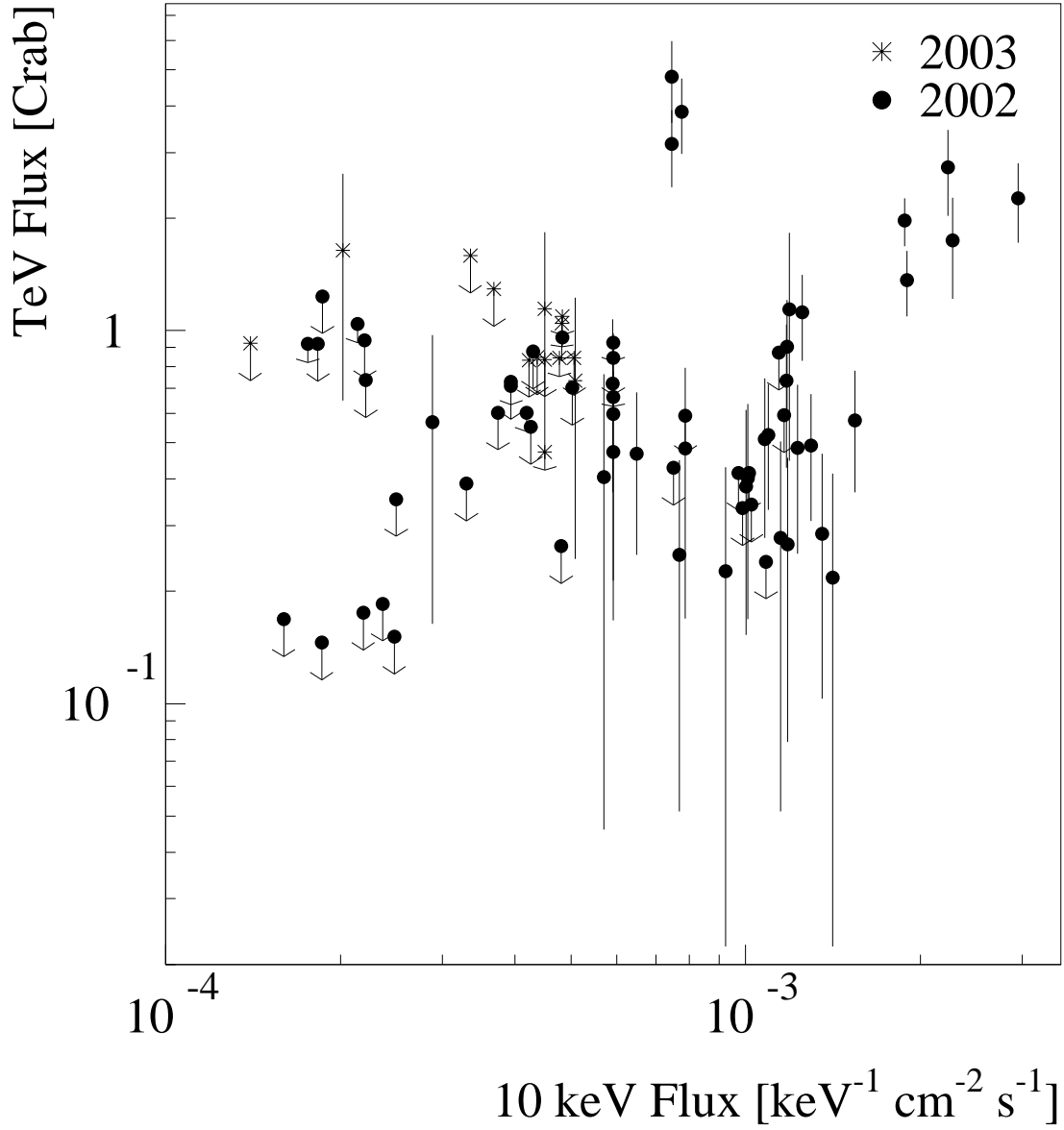}
\caption{Correlation between the X-ray and TeV $\gamma$-ray fluxes for measurements
within two hours of each other.}
\label{fig2}
\end{figure}

The correlation between simultaneously measured TeV $\gamma$-ray and X-ray fluxes 
during the 2002 and 2003 observation campaigns are shown in Figure~\ref{fig2}.  
The 2002 X-ray points are from our re-analysis of the data presented in Paper I. 
While the TeV $\gamma$-ray points are the same as in Paper I, we converted here 
all fluxes with a statistical significance below 1.6 $\sigma$ (90\% confidence) into upper limits. 
One can recognize that the main difference between the 2002 and 2003 observations 
is a lack of strong X-ray and TeV $\gamma$-ray flares during the 2003 observations. 
Since both the X-ray and TeV $\gamma$-ray fluxes were lower in 2003 than in 2002, 
the data also do not show evidence for a correlation.
\\[2ex]
The X-ray flux versus X-ray photon index correlation is shown for the 2000, 2002, 
and 2003 data sets in Figure~\ref{fig3}. The fluxes and photon indices appear to be  
correlated for the 2002 data set, and to 
a lesser extend in the 2003 data set; however, a secular evolution 
of the ``correlation'' can be recognized. For the same X-ray flux level, the 
energy spectrum was hardest during 2002, the year of the large X-ray and 
TeV $\gamma$-ray flares. We studied this effect further with a detailed spectral analysis
in three flux bands.  The June 2003 data, the entire 2000 data set, and a significant part of
the 2002 data fit within a low 10 keV flux band of 
0.1-0.4$\times10^{-3}$ cm$^{-2}$ s$^{-1}$ keV$^{-1}$.  A medium flux band was chosen
similarly and encompasses the May 2003 data and a portion of the 2002 data.  Finally, a 
high flux state represents the
flaring data in 2002.  
Details of the combined spectra in these flux bands is given in 
Table~\ref{tab3}
For the 2003 medium and 2002 high flux states the reduced 
$\chi^{2}$ values are significantly better for either a cutoff power-law
or log-parabolic model than a simple power law.  
Log-parabolic fits provide a direct way of 
calculating the peak in the total synchrotron spectrum.  
Figure~\ref{fig4} shows the 4-15 keV
spectral energy distributions from Table~\ref{tab3}  
after applying a log-parabolic fit.  
The 2003 low and medium flux state give a peak synchrotron energy of 1.21$\pm$0.64 keV and 
1.09$\pm$0.41 keV.

A similar short term relationship between
the photon index and the flux was found for Markarian 501 \citep{Kraw:00}. 
We will discuss possible explanations for this
type of secular 
evolution in Section~\ref{Gutierrez:discussion}.

\section{Spectral Energy Distribution and SSC Modeling}
\label{Gutierrez:SED}

In Figure~\ref{fig5} we show the radio to $\gamma$-ray SED of 1ES 1959+650 together 
with results from simple one-zone SSC calculations. 
The SSC code\footnote{The code can be downloaded on
\url{http://jelley.wustl.edu/multiwave/}} \cite{Kraw:04} assumes a spherical emission 
volume of radius $R$ moving with bulk Lorentz factor $\Gamma_{B}$ toward the observer.
The emission volume is filled with an isotropic electron population and a 
randomly oriented magnetic field $B$.
The energy spectrum of the electrons is assumed to be a broken power law from
energy $E_{\rm min}$ to $E_{\rm max}$ with differential spectral indices 
($p$ from $dn/dE\propto E^p$) of 2 and 3 below and above the break energy $E_{\rm b}$.
We use the CIB model of \citet{Knei:02} to calculate extragalactic absorption 
of the TeV $\gamma$-rays in pair production processes.

Figure~\ref{fig5} shows our SSC model with three different sets of parameter values,
(i) the set of model parameter values from Paper I that describes an SED observed during
2002 at intermediate flux levels, (ii) the same model parameter values adapted to fit the 2003 SED 
by reducing the $E_{\rm b}$ and $E_{\rm max}$ values; and (iii) an alternative fit
to the 2003 data allowing free variation of the parameter values.
Note that (i) and (ii) produce almost identical Inverse Compton components.
The parameter values are summarized in Table~\ref{tab6}. 
Explaining the differences between the 2002 and 2003 SED with a change of 
$E_{\rm b}$ and $E_{\rm max}$ alone results in a rather low predicted TeV 
$\gamma$-ray flux. However, both (ii) and (iii) are consistent with the data. 
We will discuss the SSC model fits further in the next section.

\section{Summary and Discussion}
\label{Gutierrez:discussion}

\begin{figure}
\plotone{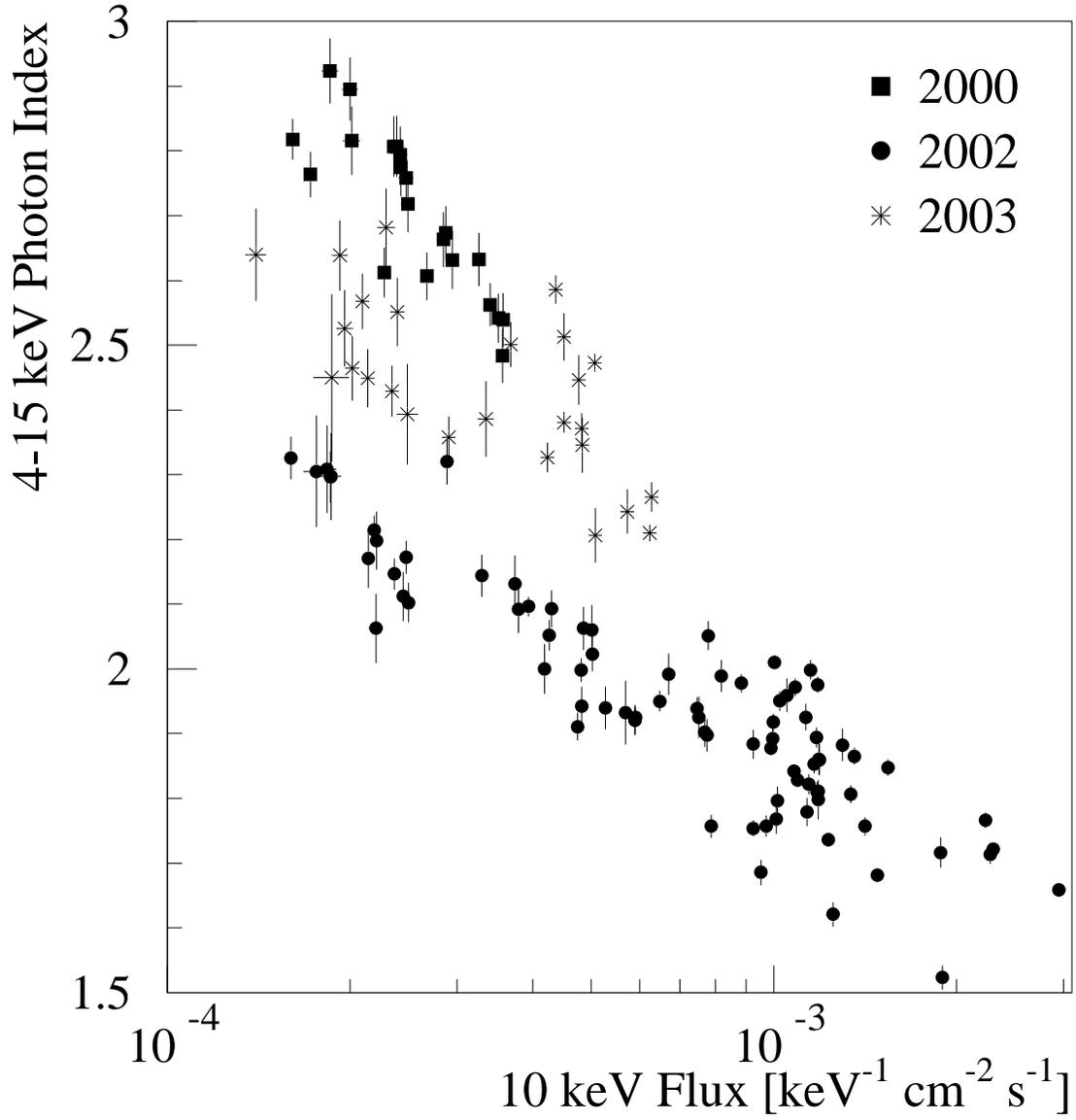}
\caption{Correlation between the X-ray flux and the 4-15 keV photon index for observations
in the years 2000, 2002 and 2003.}
\label{fig3}
\end{figure}

\begin{figure}
\plotone{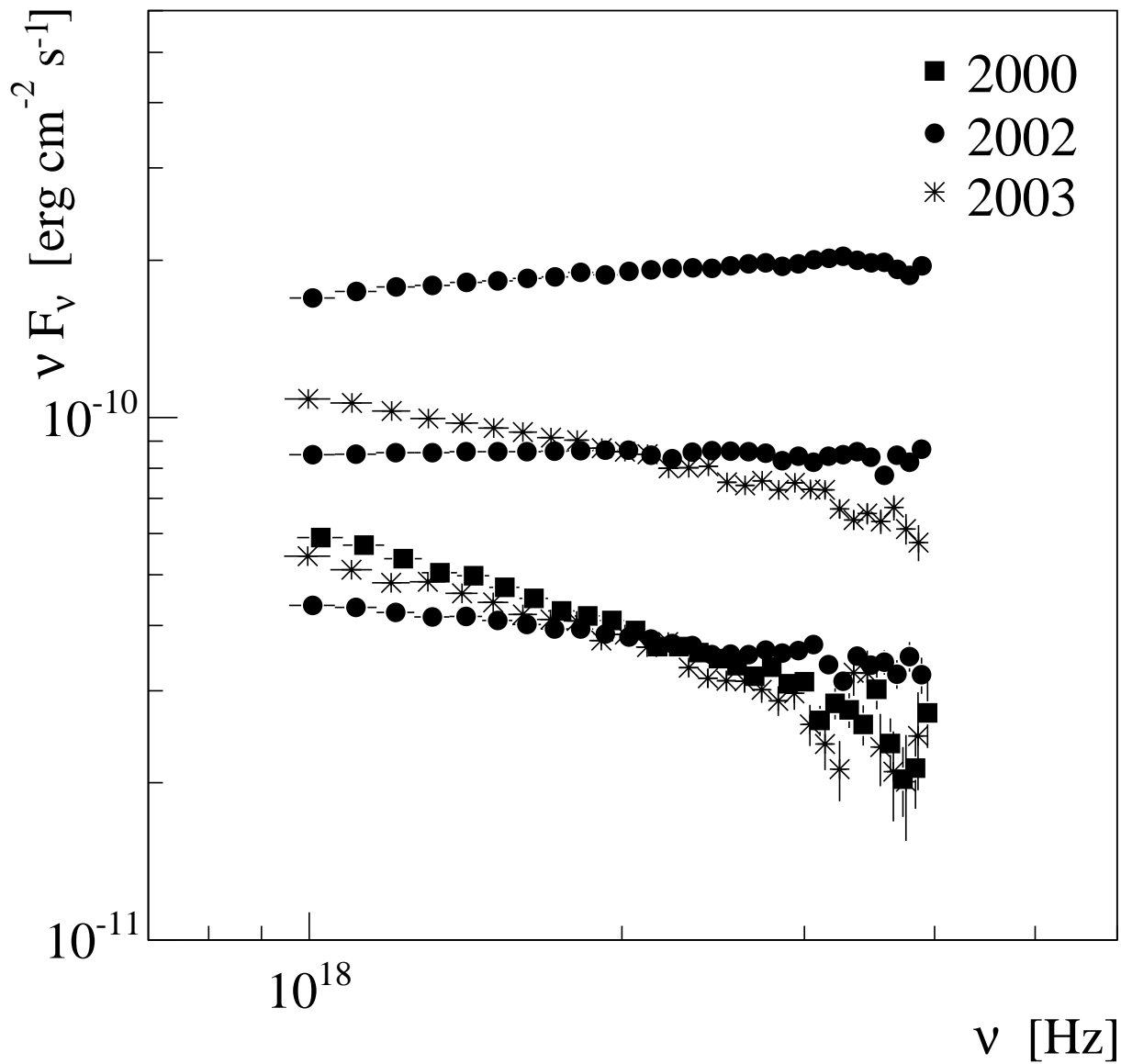}
\caption{4-15 keV X-ray spectral energy distributions from combined spectra in three flux
bands from 2000, 2002, and 2003. The plot shows for
each spectrum fitted log-parabolic model multiplied by the number of counts
detected in an energy bin divided by the number of counts expected in the bin,
given the best-fit model parameters, and the statistical errors on the data
scaled in this way. See Table~\ref{tab3} for details.}
\label{fig4}
\end{figure}

\begin{figure}
\plotone{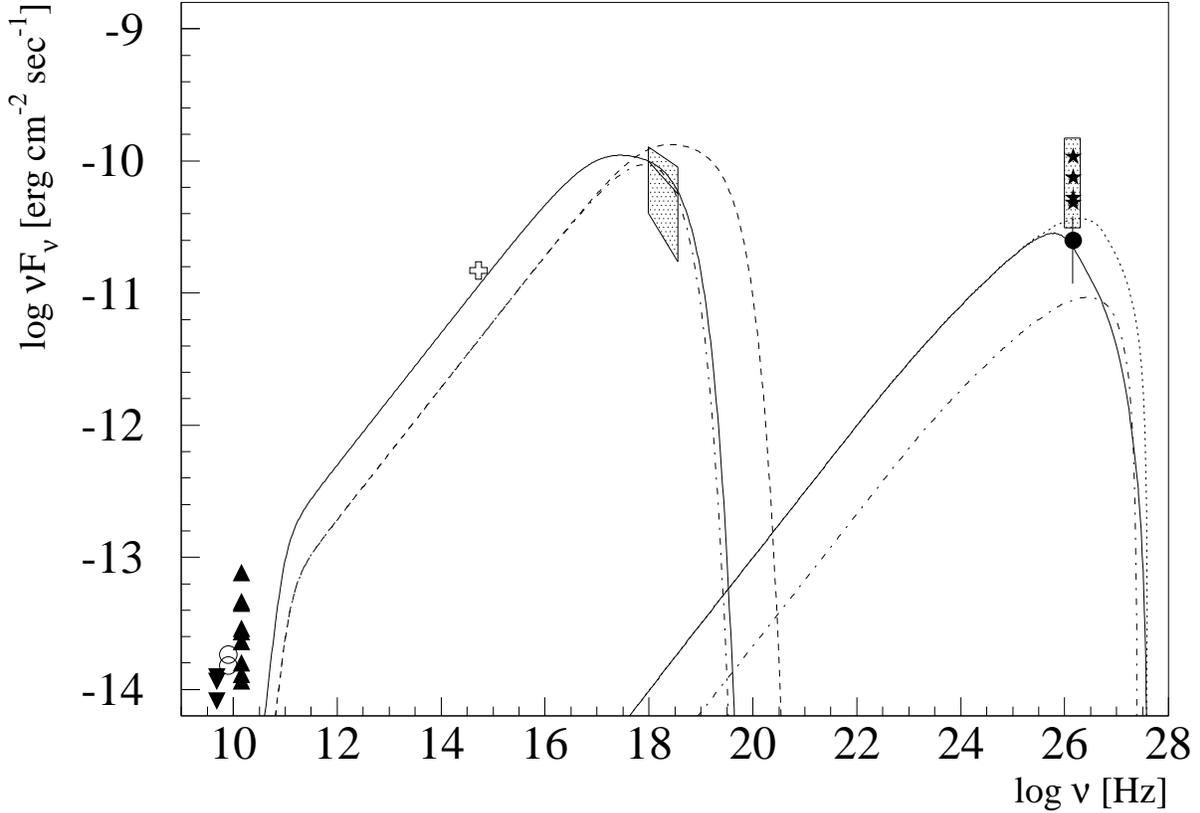}
\caption{ \small Radio to $\gamma$-ray SEDs of the blazar 1ES 1959+650.
All the radio fluxes observed between May 2, 2003 and June 7, 2003 are shown (UMRAO).
The downward pointing triangles are 4.8 GHz, the open circles are 8.0~GHz, and
the upward pointing triangles are the 14.5 GHz radio data.  
The optical data point (swiss cross) shows the flux detected on June 7, 2003 (Bordeaux Observatory).
The shaded region in the X-ray energy range shows the range of fluxes and 
energy spectra observed during the campaign with the {\it RXTE} satellite.
The straight line in the middle is the mean X-ray spectrum, measured simultaneously with the TeV 
$\gamma$-ray observation.
At TeV energies the range of upper limits are shown as a shaded region (ie. the top of the 
shaded region is the highest upper limit and the bottom of the region is the lowest upper limit)
for the majority of the 28-minute Whipple observations, and flux estimates
(solid stars) for the only four 28-minute Whipple observations which were strong enough to produce
data points (rather than upper limits).
The filled circle shows the average flux measured during the 2003 campaign in nights with RXTE observations. 
Three curves are shown: (i) the dashed line shows a SSC fit to an intermediate flux level
SED observed during 2002 (see Paper I), (ii) the dashed dotted line shows a SSC fit to the
average X-ray spectrum observed during 2003 and obtained with the same parameter values
as (i) but with a reduced break energy and high-energy-cutoff of the
electron spectrum, (iii) the solid line shows the fit to the 2003 SED with 
freely-varying SSC parameter values. All model parameter values are given in 
Table~\ref{tab6}. The dotted line shows (iii) without the effect 
of intergalactic $\gamma$-ray absorption. These three sets of model parameter values produce similar
SEDs.}
\label{fig5}
\end{figure}

\begin{figure}
\plotone{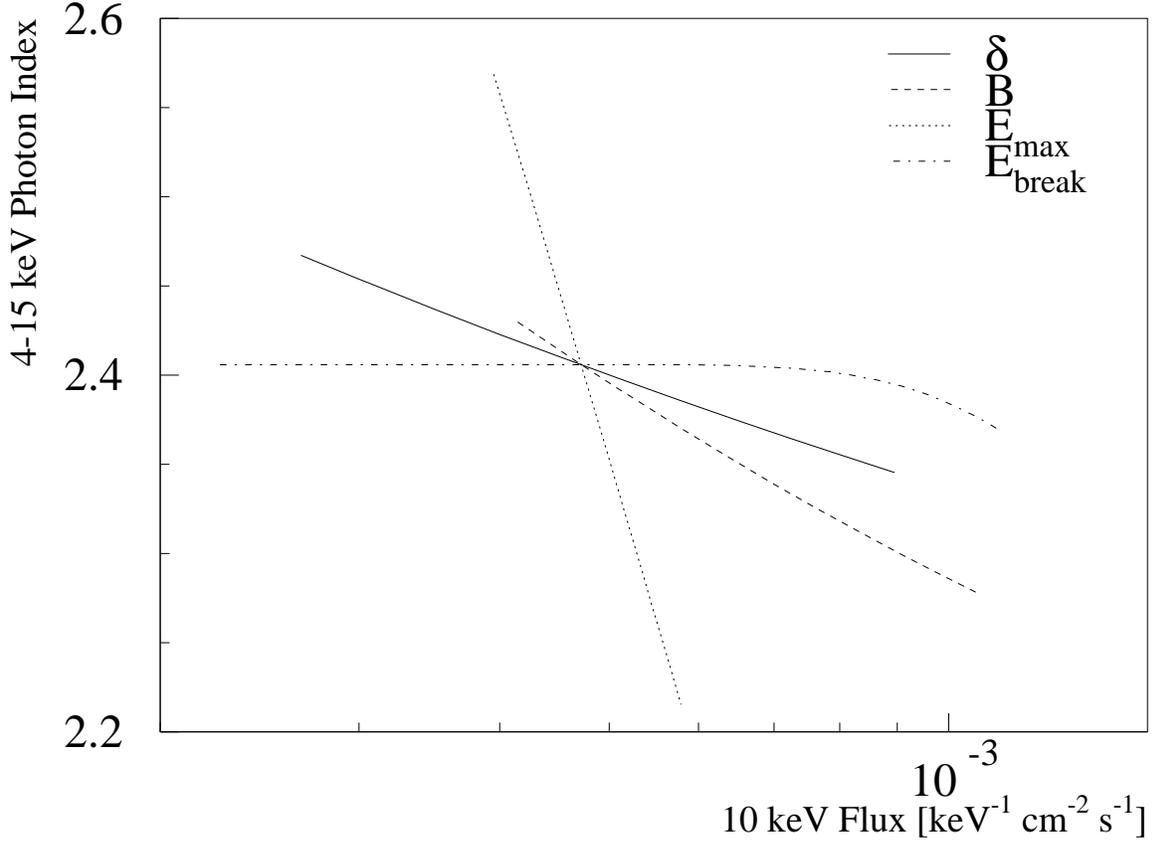}
\caption{
The 10 keV flux versus 4-15 keV photon index correlation resulting from changing
individual SSC model parameter values. All parameters not indicated are fixed to the 
values given in Table~\ref{tab6} (iii).
Solid line: $\delta$ is varied from 17.63 to 23.09 left to right;
dashed line: the magnetic field $B$ is varied from $0.019\times10^{-4}$ G to $0.028\times10^{-4}$ G from left to right;
dotted line: the parameter $\log(E_{\rm Max}/$eV$)$ is varied from 11.84 to 12.02 top to bottom;
dashed-dotted line: the parameter $\log(E_{\rm b}/$eV$)$ is varied from 10.78 to 11.47 from left to right.
The starting point for all model parameter values are given in row (iii) of Table~\ref{tab6}.
}
\label{fig6}
\end{figure}

In this paper, we have presented the results of a multiwavelength 
campaign on the blazar 1ES 1959+650 carried out in May and June 2003, 
one year after the major flaring phase of 2002. 
Our campaign did not reveal a statistically significant TeV $\gamma$-ray flare and
the highest X-ray fluxes observed during the 2003 campaign were lower by a factor 
of five than the highest fluxes observed in 2002. 
Contrasting the behavior of the source at high energies (X-rays and $\gamma$-rays),
the optical and radio flux levels did not change from 2002 to 2003.
The lack of a correlation between the high-energy (X-ray and $\gamma$-ray) 
and low-energy (radio and optical) flux levels observed in the years
2002 and 2003 suggest that the 
high-energy radiation is produced close to the central engine, and the
low-energy radiation further downstream of the jet. Indeed radio emission
produced
closer to the source
would be self-absorbed and hence not visible.  It is also possible
that optical emission may come from ``old''
components that do not shine in X-rays anymore, but could in fact be closer to
the central engine.
The examination of a possible correlation between the TeV $\gamma$-ray flux
and X-ray flux will benefit greatly from more sensitive Cherenkov telescopes such as
VERITAS or MAGIC (MAGIC detected 1ES~1959+650 in September and October 2004
in a state of low activity with a significance of 8.2 $\sigma$ after 
$\sim$ 6.0 hours of observations \citep{Albe:05}.)
\\[2ex]
An interesting result from our campaign is that the 10 keV fluxes and 4-15 keV photon 
indices appear correlated during individual observation periods, but that the 
apparent correlation evolves on a time scale on the order of a year.
In an initial attempt to identify whether a single model parameter might be responsible for
the secular evolution, we varied individual SSC model parameters and studied the effect on a 
correlation between X-ray flux and photon index.
Comparing the correlations resulting from varying individual model parameters (Figure~\ref{fig6}) 
with the apparent observed correlations (Figure~\ref{fig3}), one may speculate that individual flares
on a time scale of days are caused by a variation of the Doppler factor $\delta$, the magnetic field $B$,
and/or $E_{\rm b}$ and that
the secular evolution of the source on a time scale of months is caused by a shift of the maximum energy $E_{\rm max}$ 
to which electrons are accelerated. In a similar study of the blazar Mrk 501 (an object in 
many aspects similar to 1ES 1959+650), \citet{Tave:01} also identified
$E_{\rm max}$ as a likely parameter to cause long-term variations
of the fluxes and energy spectra. 
The results discussed here should be taken with caution, as we cannot exclude that 
several jet parameters change simultaneously from flare to flare and on longer time scales.
\\[2ex]

Recently, \citet{Uttl:05} interpreted the flaring activity of AGN as a red-noise 
process with significant power at low frequencies and a log-normal amplitude distribution. 
In this context, hourly and daily variations are considered high frequency noise,
and yearly variations low-frequency noise. 
As the discussion above shows, secular evolutions of the emission characteristics
may be able to shed light on the physical origin of the ``power at low frequencies''.
Long-term monitoring on time scales of years may thus be equally crucial for understanding 
the inner workings of AGN jets as intensive observations on time scales of several weeks.

{\it Acknowledgments:}
This research is supported by the U.S. Department of Energy, 
the National Science Foundation, the Smithsonian Institution, by 
NSERC in Canada, by Science Foundation Ireland and by PPARC in the UK.
K.~G.and H.~K gratefully acknowledge support by NASA through grant NAG 13770 
and support by the Department of Energy through the Outstanding
Junior Investigator program. The University of Michigan Radio Astronomy 
Observatory (UMRAO) is partially supported by funds from the 
Michigan Department of Astronomy.

\clearpage

\clearpage

\begin{deluxetable}{cccccccccc}
\tabletypesize{\footnotesize}
\tablewidth{0pt}
\tablecaption{Spectral fit results for the 4 keV -- 15 keV data divided 
  into three flux bands. \label{tab3}}
\tablehead{
\colhead{Year} & 
\colhead{$F_{\rm 10\,keV}\,$\tablenotemark{a}} & 
\colhead{$\chi_{Pow}^{2}$\tablenotemark{b}} &
\colhead{$\chi_{Cut}^{2}$\tablenotemark{c}} &
\colhead{$\chi_{Par}^{2}$\tablenotemark{d}} &
\colhead{$a$\tablenotemark{e}} &
\colhead{$b$\tablenotemark{f}} &
\colhead{$K$\tablenotemark{g}} &
\colhead{$E_{p}$\tablenotemark{h}} &
\colhead{$\nu_{p}F(\nu_{p})$\tablenotemark{i}} }
\startdata
2000 & 0.1-0.4 & 1.62 & 2.38 & 1.09 & 2.16 $\pm$ 0.12 & 0.28 $\pm$ 0.07 & 6.05 $\pm$ 0.63 & 0.51 $\pm$ 0.27 & 1.02 $\pm$ 0.13 \\
2002 & 0.1-0.4 & 1.12 & 1.09 & 0.97 & 2.03 $\pm$ 0.08 & 0.11 $\pm$ 0.05 & 3.16 $\pm$ 0.23 & 0.71 $\pm$ 0.62 & 0.51 $\pm$ 0.04 \\
     & 0.4-0.7 & 1.47 & 0.96 & 0.92 & 1.77 $\pm$ 0.06 & 0.13 $\pm$ 0.03 & 4.32 $\pm$ 0.24 & 6.76 $\pm$ 4.80 & 0.86 $\pm$ 0.12 \\
     & 0.7-2.5 & 2.02 & 1.03 & 0.97 & 1.62 $\pm$ 0.04 & 0.14 $\pm$ 0.02 & 7.07 $\pm$ 0.31 & 20.4 $\pm$ 13.9 & 2.00 $\pm$ 0.36 \\
2003 & 0.1-0.4 & 1.87 & 1.39 & 1.43 & 1.93 $\pm$ 0.17 & 0.38 $\pm$ 0.10 & 4.27 $\pm$ 0.63 & 1.21 $\pm$ 0.64 & 0.68 $\pm$ 0.10 \\
     & 0.4-0.7 & 2.00 & 1.02 & 1.04 & 1.98 $\pm$ 0.07 & 0.23 $\pm$ 0.04 & 8.16 $\pm$ 0.57 & 1.09 $\pm$ 0.41 & 1.30 $\pm$ 0.09 \\
\enddata
\tablenotetext{a}{\hspace*{0.2cm} 10 keV flux in units of ($10^{-3}$ photons keV$^{-1}$ cm$^{-2}$ s$^{-1}$)}
\tablenotetext{b}{\hspace*{0.2cm} Reduced $\chi^2$-value from 26 degrees of freedom for a power-law fit}
\tablenotetext{c}{\hspace*{0.2cm} Reduced $\chi^2$-value from 25 degrees of freedom for a cutoff power-law fit}
\tablenotetext{d}{\hspace*{0.2cm} Reduced $\chi^2$-value from 25 degrees of freedom for a log-parabolic fit}
\tablenotetext{e}{\hspace*{0.2cm} Photon index for a log-parabolic fit}
\tablenotetext{f}{\hspace*{0.2cm} Curvature term for a log-parabolic fit}
\tablenotetext{g}{\hspace*{0.2cm} Flux normalization for a log-parabolic fit in $10^{-2}$}
\tablenotetext{h}{\hspace*{0.2cm} Peak energy from the log-parabolic fit in units of keV}
\tablenotetext{h}{\hspace*{0.2cm} Peak flux from the log-parabolic fit in units of ($10^{-10}$ erg cm$^{-2}$ s$^{-1}$)}
\end{deluxetable}

\begin{deluxetable}{ccccccccc}
\tabletypesize{\footnotesize}
\tablewidth{0pt}
\tablecaption{Parameters values for SSC model of the in 2002 and 2003 SEDs.\label{tab6}}
\tablehead{
\colhead{} &
\colhead{$fit$} &
\colhead{$\delta\tablenotemark{a}$} & 
\colhead{$B[gauss]$} & 
\colhead{$R[m]$} &
\colhead{$w_{p}\, \tablenotemark{b}\,\,\left[ \rm erg/cm^{-3}\right]$} &
\colhead{$\log(E_{min}/$eV$)\, \tablenotemark{c}$} &
\colhead{$\log(E_{max}/$eV$)\, \tablenotemark{d}$} &
\colhead{$\log(E_{b}/$eV$)\, \tablenotemark{e}$}}
\startdata
$2002$ & $(i)$ & $20$ & $0.04\times10^{-4}$ & $14\times10^{13}$  & $0.014$ & $3.5$  & $12.2$ & $11.45$ \\
$2003\, \tablenotemark{f}$ & $(ii)$  & $20$ & $0.04\times10^{-4}$ & $14\times10^{13}$  & $0.014$ & $3.5$  & $11.70$ & $11.41$ \\
$2003$ & $(iii)$ & $20$ & $0.02\times10^{-4}$ & $27.2\times10^{13}$ & $0.010$ & $3.5$  & $11.90$ & $11.10$  \\
\enddata
\tablenotetext{a}{\hspace*{0.2cm} Doppler factor of plasma}
\tablenotetext{b}{\hspace*{0.2cm} Electron energy density in $erg/cm^{3}$}
\tablenotetext{c}{\hspace*{0.2cm} Minimum energy of electrons}
\tablenotetext{d}{\hspace*{0.2cm} Maximum energy of electrons}
\tablenotetext{e}{\hspace*{0.2cm} Energy where electrons spectrum index changes from a 2 to 3}
\tablenotetext{f}{\hspace*{0.2cm} Best fit for 2003 Data with only $\log(E_{max}/$eV$)$ and $\log(E_{b}/$eV$)$ varied from 2002.}
\end{deluxetable}
\end{document}